# OMNIRank：基于深度学习的 P2P 平台风险量化研究

张宏伦 [1] 王海洋 [1] 陈夏明 [1] 王永坤 [2] 金耀辉 [1, 2]

1（上海交通大学 区域光纤通信网与新型光通信系统国家重点实验室，上海 200240）

2（上海交通大学 网络信息中心，上海 200240）

（jinyh@mail.sjtu.edu.cn）

# OMNIRank: Risk Quantification for P2P Platforms with Deep Learning

Honglun Zhang[1] Haiyang Wang[1] Xiaming Chen[1] Yongkun Wang[2] Yaohui Jin[1,2]

1 (State Key Lab of Advanced Optical Communication System and Network, Shanghai Jiao Tong University, Shanghai, 200240)

2 (Network and Information Center, Shanghai Jiao Tong University, Shanghai, 200240)

**Abstract** P2P lending presents as an innovative and flexible alternative for conventional lending institutions like banks, where lenders and borrowers directly make transactions and benefit each other without complicated verifications. However, due to lack of specialized laws, delegated monitoring and effective managements, P2P platforms may spawn potential risks, such as withdraw failures, investigation involvements and even runaway bosses, which cause great losses to lenders and are especially serious and notorious in China. Although there are abundant public information and data available on the Internet related to P2P platforms, challenges of multi-sourcing and heterogeneity matter. In this paper, we promote a novel deep learning model, OMNIRank, which comprehends multi-dimensional features of P2P platforms for risk quantification and produces scores for ranking. We first construct a large-scale flexible crawling framework and obtain great amounts of multi-source heterogeneous data of domestic P2P platforms since 2007 from the Internet. Purifications like duplication and noise removal, null handing, format unification and fusion are applied to improve data qualities. Then we extract deep features of P2P platforms via text comprehension, topic modeling, knowledge graph and sentiment analysis, which are delivered as inputs to OMNIRank, a deep learning model for risk quantification of P2P platforms. Finally, according to rankings generated by OMNIRank, we conduct flourish data visualizations and interactions, providing lenders with comprehensive information supports, decision suggestions and safety guarantees.

**摘要** 相对于银行等传统机构，P2P 网络借贷的出现带来了一种新颖且灵活的借贷形式。在 P2P 网贷中，无需复杂的审核流程，投资者和借贷人便可直接完成资金对接并互惠互利。然而由于缺乏相关法律、委托监管和有效管理，P2P 平台存在提现失败、经侦介入和跑路等潜在风险，有可能给投资者造成极大损失，并且这一问题在中国尤为严重。互联网上虽然有大量和 P2P 平台相关的公开资讯和数据，但存在多源异构、质量不齐等挑战。针对这一需求，本文提出一种基于深度学习的 P2P 平台风险量化模型，融合 P2P 平台的多维特征并进行评分排名。研究首先搭建了一套大规模分布式和灵活可扩展的爬虫框架，从互联网上获取了国内 P2P 平台从 2007 年至今的海量多源异构数据，并采用数据去重、空值处理、数据去噪、格式统一、对齐融合等清洗提升数据质量；

然后通过文本理解、主题模型、知识图谱、情感分析等方法，逐步深入地提取 P2P 平台的多维特征，并提出了基于深度学习的 P2P 平台风险量化模型 OMNIRank；最后根据 OMNIRank 的排名结果，进行了丰富的数据可视化和交互探索，为广大投资者提供全面的信息支持、决策建议和安全保障。

P2P（Peer-to-Peer，即个人到个人）网络借贷于 2007 年随互联网浪潮引入国内，2013 年开始蓬勃发展，平台数量和交易金额都呈现大幅增长。相对于银行等传统金融借贷机构，P2P 网贷具有门槛低、放款快、审核容易、灵活性强等优势，投资者和借贷人可以直接完成资金对接并互惠互利，由于信用问题无法通过银行审批的借贷人仍有机会获得借款，而投资者往往可以享受到比银行存款更高的利润。

P2P 平台是一个个相对独立的 P2P 网贷运营公司。由于目前国内缺乏针对 P2P 网贷制定的相关法律、专有部门的委托监管和成熟有效的管理方法，P2P 平台存在提现失败、经侦介入、跑路倒闭等潜在风险。截止至 2016 年 5 月，国内 P2P 平台共上线 4080 家，其中 1307 家出现问题，给投资者造成巨大损失的同时，也严重影响了国内互联网金融的行业氛围。举例来说，影响最为恶劣的 e 租宝公司，在短短一年内便非法集资 500 多亿，涉及投资者 90.95 万人。

投资者关注和了解 P2P 平台的主要途径是通过网络。互联网上虽然有大量和 P2P 平台相关的公开资讯和数据，例如新闻报道、网民评论、平台资料和交易数据，但呈现出多源异构、质量参差不齐、存在缺失和错误等一系列挑战。这些公开网络数据需要经过恰当的处理和融合，才能对 P2P 平台的潜在风险形成较为全面准确的风险评估。

针对以上问题，本文提出一种基于深度学习的 P2P 平台风险量化模型 OMNIRank，该模型基于和 P2P 平台相关的各类公开网络数据，在经过数据获取、清洗和分析后，提取出能够反映 P2P 平台风险信息的多维特征，并以此为输入进行全面精确的风险量化。结果显示，OMNIRank 能达到 86%以上的预测准确率，并为投资者筛选出一个零风险投资区间。根据 OMNIRank 的排名结果，本文进行了丰富的数据可视化和交互探索，在平台风险得到保障的前提下，辅助投资者选择适合自己的投资平台。本文的系统架构如图 1 所示，主要解决了以下两个问题：

1) **公开网络数据的获取、处理和分析**。互联网上的公开数据具有多源异构、数量巨大、非结构化、质量不齐等问题，举例来说，和 P2P 平台相关的新闻报道可能来自多个门户，各个门户的用词方式、报道角度不尽相同，网民对于 P2P 平台的观点和评论也没有统一的格式规范。除此之外，以上文本数据和 P2P 平台的实时交易数据都在不断产生和积累，对这些海量数据进行获取、处理和分析，并从中提取出和 P2P 平台风险有关的信息，具有很大的挑战；

2) **基于多维特征的 P2P 平台风险量化**。P2P 平台风险是一个综合指标，和多个方面的因素和指标都存在潜在关联，例如平台的注册信息、负面新闻和舆论、平台数据异常等。这些特征可能对应一个值或者序列，数据类型也可能是数值、类别值或者文本。如果不能系统地将这些特征进行融合，很有可能得到局部片面和不准确的量化结果。

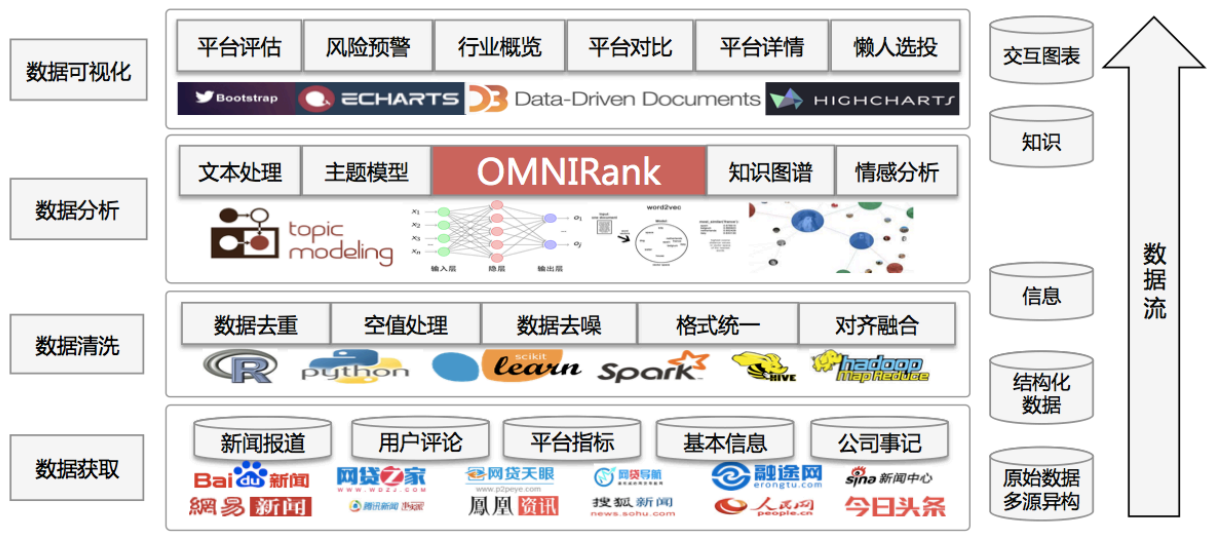

Fig.1 The framework of Risk Quantification

图 1 P2P 平台风险量化总体架构图

在下文中，第一、二节分别介绍相关研究工作和本文中数据源的选取，第三、四节讨论海量公开网络数据的获取、清洗和分析，从多源异构数据中提取 P2P 平台的多维特征，第五节阐述平台风险量化模型 OMNIRank 以及其性能评估，第六节介绍在线可视化和交互产品，辅助投资者选择适合自己的平台。全部研究成果都已开源，详细链接参见附录。

## 1 相关工作

P2P 网贷的发展吸引了大量学术界的研究工作，主要可以分为以下三大类：

1) **投资者和借款人的行为模式**。将 P2P 平台理解为社交网络和金融借贷的组合，可以将社交网络领域的相关研究成果应用到 P2P 网贷中，如网络的图模型，网络的产生、发展和演化过程登；研究不同信用等级借款人的统计特征，以及投资者的跟风行为和分散投资策略等[1]；
2) **影响借款成功与否的可能因素**。从信用特征和社交特征[2]两个角度研究影响借款交易的可能因素，前者是指个人信用记录、信用卡消费记录等个人特征，后者是指借款人在 P2P 平台上的人脉关系、群组认证、他人评价等社交特征；
3) **辅助投资者进行最优投资**。一方面建立风险模型，根据借款人各项特征判断交易是否存在风险；另一方面综合风险和回报，考虑两者之间的关联和影响，在风险阈值内追求最高回报[3]。

可以看出，已有研究大多只是从微观角度研究单个投资人、借款人或单笔借款交易的特征和性质，未能从宏观和长期的角度对 P2P 平台整体进行持续准确的风险量化和监控，而后者对于投资者的资金安全具备更为显著的研究意义。

另外，国内已有很多较为活跃的 P2P 网贷社区，例如网贷之家[1]、网贷天眼[2]、融 360[3]等，但这些网贷社区只是简单地收集新闻报道、网民评论、平台资料和数据等内容并存储和展示，而未进行深入的挖掘和融合，投资者面对的仍然是不同类型的大量原始数据，并不能快速直观地获取最为核心的平台风险信息。

本文从宏观和长期的角度出发，从海量多源异构数据中挖掘 P2P 平台的多维特征，基于深度学习对平台本身进行持续全面的风险量化，因而具备更高的实际应用参考价值。

## 2 数据源选择

在数据源的选择上，应充分涵盖和 P2P 平台风险信息有关的各类数据[4]。在经过大量的调研之后，本文将 P2P 平台出现问题的原因总结为以下两方面：先天基因存在不足和后天发展出现问题。前者包括注册资金不足、创始人团队信用不可靠、所处城市不利于行业发展等，这些因素在平台注册成立时便已确定，并将一直影响平台的后期发展；后者包括过度鼓吹利率而无法兑现、风控不过关造成大量逾期、缺乏投资者导致资金链断裂、金钱利益诱发的违法犯罪等，此类原因是在平台的发展过程中出现的，具有动态性和不确定性等特点。将这些原因对应到具体的平台特征，主要可以分为以下四大类：

1) **静态特征**，包括平台的性质和类别、平台标签、上线时间、注册资金、所在城市、投标保障、保障模式、担保机构、人员信息等；
2) **动态指数**，包括平台成交量、利率、历史待还、资金净流入、投资人数、借款人数、借款标数、平均借款期限、标的金额分布等；
3) **动态新闻**，包括新闻报道内容、正面新闻数量、负面新闻数量、不同主题新闻分布、平台口碑、重点事件、平台关键词、平台语义等；
4) **动态评论**，包括用户评论内容、正面评论数量、负面评论数量、用户画像、舆情标签、评论关键词、舆情倾向、主流观点等。

以上特征从不同的角度直接或间接地反映出平台的风险信息，需要进行系统全面的挖掘融合。然而第一类特征是平台的固有属性，一般来说很少改变，因此可以用单个值来表示，而后三类特征在平台的发展过程中每时每刻都不断更新，因此对应一个值序列。同时，特征的类型可以是数值、类别值或者文本，传统的机器学习模型大多无法处理复杂异构的数据输入，而 OMNIRank 通过使用多种深度学习模型组成的复杂网络来解决这一问题。

根据以上讨论，本文选取了一些国内公开网站作为数据源，包括主流新闻门户和社交媒体、活跃的网贷社区和各大 P2P 平台的官方网站，从尽可能多的维度来挖掘 P2P 平台的风险信息。数据源的详细介绍如表 1 所示。

## 3 数据获取和清洗

为了从以上提及的众多数据源中获取海量多源异构数据，本文基于开源爬虫框架 Scrapy 进行开发和改进，并集成了 Redis 和 Hadoop。Redis 是性能优越的缓存数据库，可以用于服务器之间的同步和多任务

1 网贷之家，http://www.wdzj.com/
2 网贷天眼，http://www.p2peye.com/
3 融 360，http://www.rong360.com/

Table 1 Overview of data sources

表 1 数据源概览

| 描述 | 网站 | 作用 |
|---|---|---|
| 新闻门户 | 网易新闻、凤凰资讯、搜狐新闻、腾讯新闻、人民网、新浪新闻 | 原始动态新闻 |
| 网贷社区 | 网贷之家、网贷天眼、网贷导航、融 360 | 平台资料和动态指数 |
| 社交媒体 | 新浪微博、百度贴吧、知乎、豆瓣、天涯社区 | 原始动态评论 |
| 平台官网 | 拍拍贷、陆金所、人人贷、宜人贷、点融网 | 官方数据和平台大事 |

之间的调度，而 Hadoop 是一个分布式系统基础架构，可用于海量数据的存储和管理。除了大规模分布式计算能力外，本文还实现了灵活可扩展的配置模块，使得对于添加或者更新爬取任务，只需要简单地修改配置即可。

对于第二节中讨论的众多数据源和数据获取任务，经过简单的定制、为每项获取任务配置相应的主页域名、解析规则和目标字段，通过调度、爬取、解析、存储四大模块的协作，即可在短时间内快速获取海量多源异构数据。表 2 总结了本文所获取的部分数据资产，包括官方新闻、网民评论、平台资料、人员信息、评级数据、行业指数、地域统计和类别统计等，具有较强的异构性。在所获取的 3050 家 P2P 平台中，1672 家仍在正常运行，其他 1378 家则由于提现失败、经侦介入或跑路等问题而停业，并且发生问题的时间已知，因此可以将平台是否正常运行作为之后模型训练的标签。

获取了以上多源异构的数据资产之后，本文进行了必要的清洗工作以提升数据质量[5]，包括去除重复和类似的新闻报道、将空值更改为对应的默认值、使用 UGC 算法去除低质量网民评论、使用基于密度的聚类去除无关新闻报道、将多源异构数据转换为统一的数据表达形式、汇聚多源异构数据中的一致部分等。以 UGC（User Generated Content）算法为例，本文使用以下模型对网民评论内容进行评分并去除评分低于 0.2 的记录：

$$UGC_i = N(T_i \times 5 + E_i \times 3 + W_i \times 2)$$

其中$T_i$表示评论$i$的 TfIdf 得分，$E_i$表示其情感得分，即态度倾向是否明显，$W_i$表示评论的用户权重，即该网民的发言次数占总次数的百分比，$N$为归一化

Table 2 Overview of data capital

表 2 所获数据资产概览

| 数据集名称 | 描述 | 数值类型 |
|---|---|---|
| 官方新闻 | 来自 1908 个新闻门户的 270815 条新闻报道 | 文本序列 |
| 网民评论 | 86951 条和 P2P 网贷相关的网民评论 | 文本序列 |
| 平台资料 | 3050 家 P2P 平台的基本信息、核心指数、历史数据和大事件 | 数值、类别值、文本、数值序列、文本序列 |
| 人员信息 | 6512 名平台高官个人信息 | 文本 |
| 评级数据 | 从 2013.8 至 2016.4 所有平台的评级指数数据 | 数值序列 |
| P2P 行业指数 | 从 2013.8 至 2016.4 全部平台和问题平台数量、历史交易指数和人气指数 | 数值序列 |
| 地域统计 | 不同地域平台的历史数据统计，如北上广 | 数值序列 |
| 类别统计 | 不同类型平台的历史数据统计，如民营、银行、风投、国资 | 数值序列 |

函数，将所有评论的 UGC 值映射到 0~1 区间中。表 3 给出了 UGC 算法的部分评分结果，可以看出，得分越高的评论所含信息量越多。

Table 3 Examples of UGC scores

表 3 UGC 评分结果示例

| 评论内容 | UGC 评分 |
|---|---|
| 人气旺，一天 24 小时发标，但难抢，资金不站岗，保护投资人利益，体验好。社区也热闹。 | 0.8835 |
| 今天上线的 1 个月的周转贷没有抢到，请问后期还会推出短期项目吗？大概什么时候？ | 0.3804 |
| 降息太快了。 | 0.1765 |
| 感觉可长靠。 | 0.1333 |

## 4 数据分析

数据分析的目的是从以上数据资产中提取出之前所讨论的 P2P 平台四大类特征，从不同的维度刻画和反映出平台所隐含的风险信息。平台的静态特征和动态指数可以直接从平台资料数据集中获得，而动态新闻特征和动态评论特征则需要使用自然语言处理技术进行提取。通过以下文本理解、主题模型、知识图谱、情感分析四个步骤，本文提取出了静态特征、动态指数、动态新闻、动态评论中的绝大多数平台风险特征，这些特征将作为 OMNIRank 模型的输入，为 P2P 平台风险量化提供全面丰富的特征基础。

## 4.1 文本理解

本文首先使用开源中文处理组件结巴分词对原始新闻报道和网民评论进行了初步处理，包括中文分词、移除停用词、词性标注、关键词提取等。采用 Word2Vec 模型基于全部的新闻和评论语料进行 Word Embedding，将语料中的词语训练成 256 维词向量。在词向量的基础上，可以通过余弦相似度、欧几里德距离等方法计算词语之间的相似度，从而赋予词语更加丰富的语义信息。

## 4.2 主题模型

在理解文本的基础上，采用 Latent Dirichlet Allocation 主题模型[6]对新闻进行聚类。LDA 属于无监督的文本分类模型，输入为文档-词语矩阵，矩阵中的每个元素表示词语在文档中出现的次数，通过矩阵分解输出两个矩阵：文档-主题矩阵、主题-词语矩阵，使得同一主题的文档所使用的词语尽可能相似。本文将聚类数量设为 5，经过聚类之后，每个主题中出现概率最大的 7 个关键词如下所示：

- 主题 1（行业）：市场、配资、规模、指数、需求、房地产、理财产品；
- 主题 2（观点）：表示、合作、未来、可以、没有、成为、模式；
- 主题 3（政策）：中国、管理、政府、央行、出台、支持、银监会；
- 主题 4（平台）：公司、担保、融资、项目、有限公司、抵押、典当行；
- 主题 5（理财）：投资者、投资人、资金、P2P、平台、数据、收益。

根据这些关键词可以大致推断每类所对应的主题，分别为行业、观点、政策、平台、理财，分别涉及 P2P 网贷的不同关注领域。各个平台在不同主题上的新闻数量分布规律和变化趋势可用于事件检测和平台分析，这也将作为平台的动态新闻特征输入到 OMNIRank 模型中。

## 4.3 知识图谱

知识图谱是一种知识表示的形式，用节点表示实体、用有向边或无向边表示实体之间的关联，因此可以用图论中的相关理论来研究实体的性质，使用最短路径等算法发现实体之间的关联。同时节点和边都可以具备丰富的属性，从而能够更完善地组织和表示现实世界中的知识。

在对文本语料进行命名实体识别、关系抽取、开放域知识提取等处理后，本文构建了一张包含 3050 家平台、6512 名人员、1680 种职位、15 种平台标签、8 类平台性质、29 处地理分布，共计 11294 个节点的知识图谱，并将其存储于图数据库 Neo4j 中以实现更快的查询速度。节点之间可能存在多种关联，例如，根据所获取的平台数据，平台节点会和对应的人员、平台标签、平台性质、地理分布等多个节点存在关联。

在知识图谱的基础上，可以探索平台之间的关联和相似度，和平台直接关联的节点特征也隐含了平台的风险信息。举例来说，标签、性质和地理分布都相同的平台之间可能存在更大的相似度，而平台如果出现基本信息缺失的情况，即没有与相应类别的节点关联，或者和已知的问题平台相似，都有可能存在更大的风险。

## 4.4 情感分析

情感分析（Sentiment Analysis）是指发现和挖掘文本中所包含的正负情感词和隐含语义等内容，判断文本所表达的情感倾向性和用户态度。通过对新闻报道和网民评论进行情感分析，可以获悉媒体和公众对 P2P 平台所持态度，强烈的负面态度可能意味着平台风险的增加。

本文基于互联网买家评论数据集，训练了一个半监督递归自编码器模型用于情感分析[7]，将所获取的 270815 条新闻和 86951 条评论分为正面和负面两类，并通过 NVIDIA GPU 加快模型训练和预测速度，从而得到各个平台以天为粒度的正面以及负面新闻和评论数量，即动态新闻和动态评论中与正负情感有关的特征。

# 5 P2P 平台风险量化

给定 P2P 平台的特征集合$\vec{X}$，包括静态特征集$\overrightarrow{X_s}$、动态指数特征集$\overrightarrow{X_{di}}$、动态新闻特征集$\overrightarrow{X_{dn}}$、动态评论特征集$\overrightarrow{X_{dc}}$，即$\vec{X}=\overrightarrow{X_s}\cup\overrightarrow{X_{di}}\cup\overrightarrow{X_{dn}}\cup\overrightarrow{X_{dc}}$，同时给定各个平台的风险标签$L\in\{0,1\}$，1 表示平台正常运营，0 表示为问题平台。平台风险量化的目标是得到以下评分函数，以特征集合$\vec{X}$为输入并输出风险评分$S$，其中$S\in[0,1]$。

$$S = f(\vec{X})$$

风险量化值$S$应当满足以下三点性质：

1. 选定某个合适的阈值$\beta$后，$S$值大于$\beta$的应尽可能为正常平台，小于$\beta$的应尽可能为问题平台；
2. 正常平台的$S$值应尽可能高并接近 1，而问题平台的$S$值应尽可能低并接近 0；
3. 将所有平台按$S$值从高到底排列后，越靠前的平台中存在的问题平台应尽可能越少。

因此，P2P 平台风险量化本质上是一个有监督的二分类问题[8]，以上提出的三点性质可以作为模型性能的评估标准。为了从多维异构的平台特征中全面准确地融合出风险信息并进行评分，本文提出了基于深度学习的 OMNIRank 模型。

深度学习在近几年的研究中得到了蓬勃的发展和广泛的关注[9]，随着 GPU 性能的提升实现了进一步的普及，在自然语言处理、模式识别、语音识别和信息处理等领域表现尤为卓越。深度学习模拟人脑的思维模式，能够通过深层网络抽象出更为高层的概念并进行决策，而这正是投资者从纷繁复杂的数据海洋中概括有用信息所需要的能力。OMNIRank 采用多种深度学习模型组合成的神经网络来处理不同形式（单个值或值序列）和不同类型（数值、类别值、文本）的输入特征，进行全面精确的 P2P 平台风险量化并生成评分和排名。数值特征可以直接输入到 OMNIRank 中，类别值特征需要经过 One-Hot 处理转换成 0-1 向量，文本特征则通过 Word2Vec 模型转换成词向量并输入。

## 5.1 OMNIRank

OMNIRank 的模型结构如图 2 所示，共包含 5 类特征输入，分别对应静态数值特征、静态类别特征、动态指数、动态新闻和动态评论。本文采用卷积神经网络（CNN，Convolutional Neural Network）和最大池化层（Max-Pooling）处理静态类别特征，使用全连接层（Dense）处理静态数值特征；动态指数特征以序列的形式输入到长短时记忆（LSTM，Long Short-Term Memory）中，而动态新闻和动态评论特征则输入到 LSTM 和 CNN 的并行组合中；以上网络层的输出传递到隐藏层中进一步融合和汇聚，并加入 Dropout 层避免过拟合和提高模型泛化能力。

除此之外，本文还考虑了不同类别输入特征之间可能存在的相似关联和相互影响，并将两者组合到同

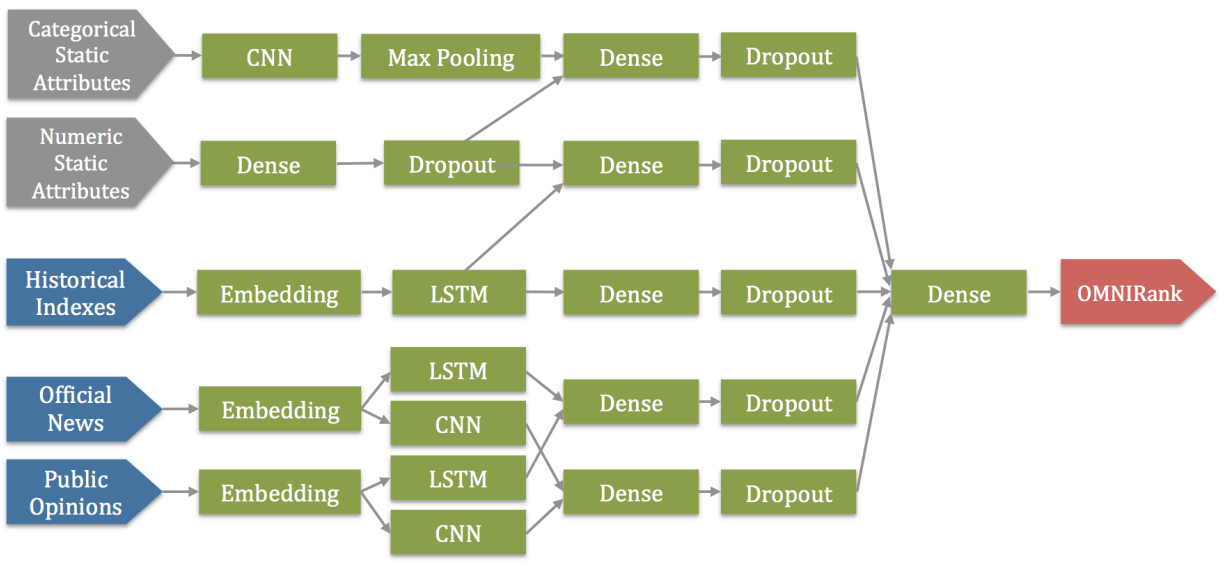

Fig.2 The framework of OMNIRank

图 2 OMNIRank 模型结构图

一个全连接层中，例如都属于平台固有属性的静态数值特征和静态类别特征，同样是数值类型的静态数值特征和动态指数特征，以及同样是文本类型的动态新闻特征和动态评论特征。通过以上结构，OMNIRank 能够综合 P2P 平台的多维异构特征，基于尽可能多的数据输入从宏观和长期的角度进行平台风险量化。

## 5.2 模型性能评估

本文使用基于 Theano 和 TensorFlow 的开源深度学习组件 Keras 实现了 OMNIRank 模型，采用 5 分交叉验证进行模型训练和预测。由于在 P2P 平台风险量化这一问题上暂无相关研究和模型可作为比较，本文选择了支持向量机（SVM，Support Vector Machine）、随机森林（RF，Random Forests）和逻辑回归（LR，Logistic Regression）三种经典机器学习分类模型来比较 OMNIRank 的性能。

根据性质 1，由于问题平台以及问题出现的时间都为已知，本文选择每个月为时间节点，使用以上四种模型（OMNIRank、SVM、RF、LR）进行风险量化和评分，并将评分排名中前 60%的平台标记为正常平台，而后 40%的平台标记为问题平台。在每次评分时，仅使用当下时间以前的输入特征数据，而排除当下时间以后的输入特征数据，各个平台的标签同样结合当下时间和问题出现时间进行判断和更新。相对于使用全部输入特征数据和各个平台的最新状态作为标签，以上方法可以使得每个月的评分结果更加符合实际。四种模型的标记准确率如图 3 所示，可以看出从 2015 年 11 月至 2016 年 4 月的 6 次测评中，OMNIRank 始终保持最高的准确率，并且随着数据量的积累和增长，准确率整体呈现上升趋势，在最近的一次评分中已经达到了 85%的准确率。

考虑到风险量化的最终目标是识别出问题平台，本文也使用了 AUC 作为评估指标来比较以上四种模

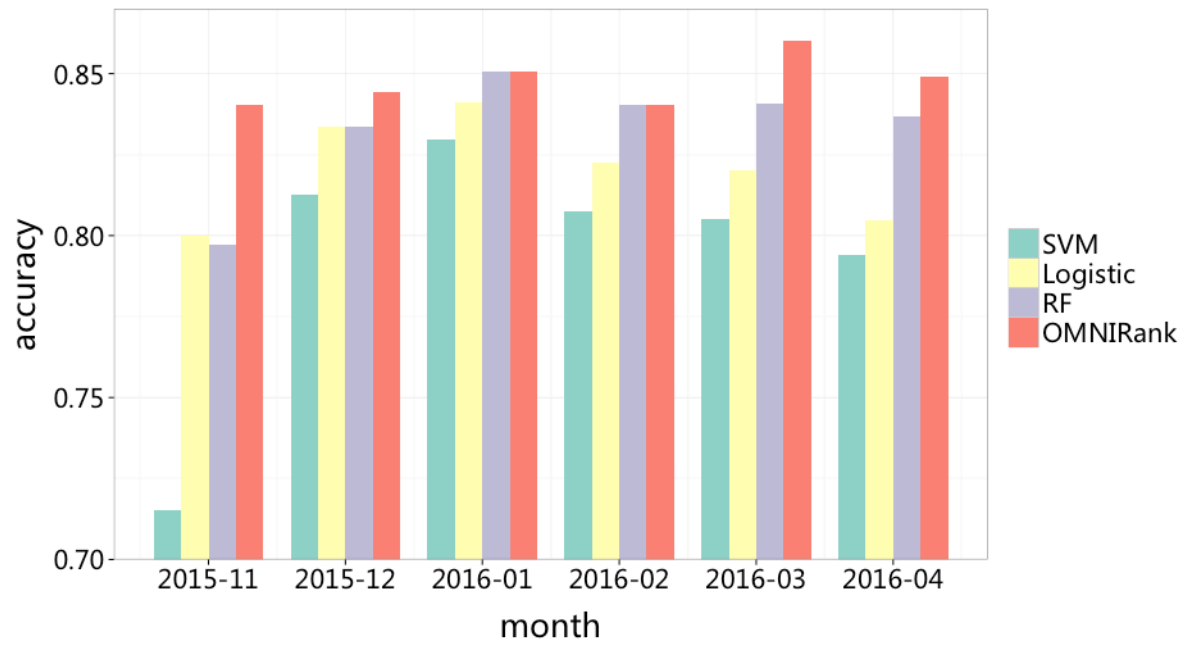

Fig.3 Comparisons of models' accuracies

图 3 模型准确率比较

型的性能。AUC 值的定义如下：

$$AUC = \frac{\sum_i S_i}{M \times N}$$

其中$M$和$N$分别为正例（正常平台）和反例（问题平台）的数量。对于每一个正例-反例对（共计$M \times N$对），假设模型对正例的评分为$s_{ip}$，对反例的评分为$s_{in}$，则该正例-反例对的得分$S_i$根据$s_{ip}$和$s_{in}$的相对大小关系决定：

$$S_i = \begin{cases} 1 & s_{ip} > s_{in} \\ 0.5 & s_{ip} = s_{in} \\ 0 & s_{ip} < s_{in} \end{cases}$$

不难看出，AUC 值位于 0 和 1 之间且越高越好，更高的 AUC 值意味着模型给正例的评分整体相对于反例更高，而不需要一个绝对的阈值用于划分正例和反例，因此更适合于风险量化模型性能的评估。四种模型的 AUC 值如图 4 所示，OMNIRank 的性能明显优于 SVM 和 LR。尽管 RF 在 2015 年 12 月的评分中 AUC 值略微高于 OMNIRank，但随着数据量的积累和增加，OMNIRank 拥有更强的学习能力、AUC 值提升更快并超过了 RF，在最近一次评分中达到了 0.91 的 AUC。

根据性质 2，风险量化的目标是给正常平台尽可能高的评分，给问题平台尽可能低的评分，从而便于将两者进行区分。图 5 显示的是在四种模型最近一个月的评分结果中，正常平台和问题平台在不同评分区间的数量分布。可以看出，OMNIRank 明显地将正常平台和问题平台区分开来，并且在 1 和 0 处分别出现了两类平台所对应的峰值；相较之下，RF 的结果中，正常平台的评分不够高，且不少问题平台得到了较高的评分；LR 和 SVM 则表现更差，相当数量的问题平

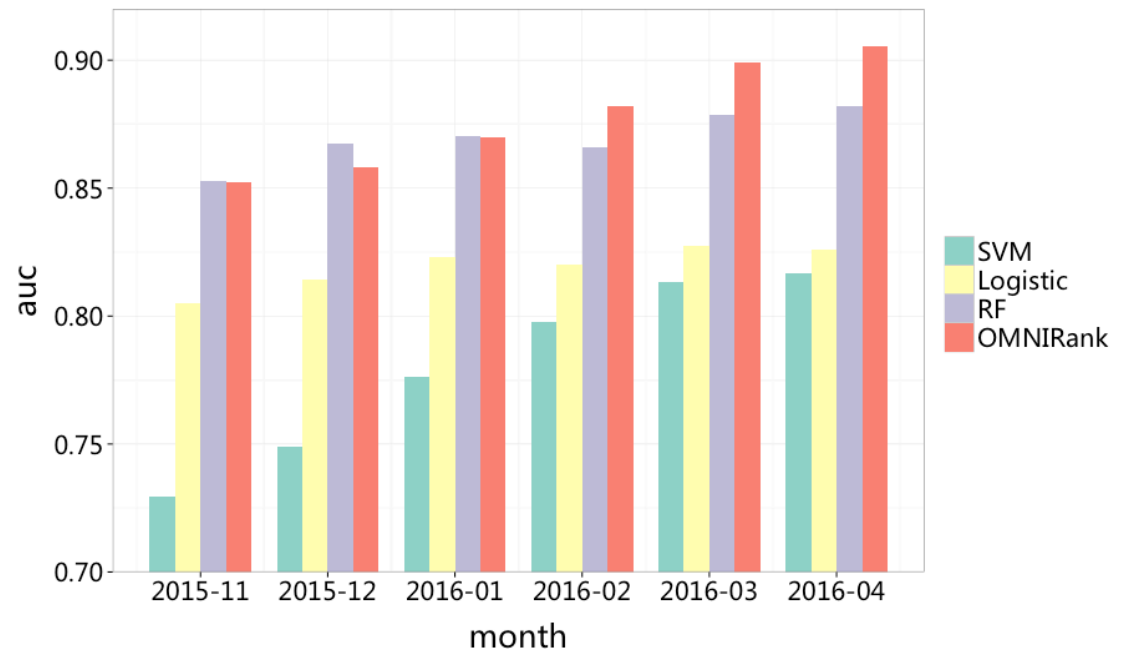

Fig.4 Comparisons of models' AUC

图 4 模型 AUC 值比较

台得分和正常平台混叠在一起（0.2 和 0.8 附近），而且 1 和 0 两端也没有出现显著峰值，未能明显地将正常平台和问题平台划分到不同的评分区间中。

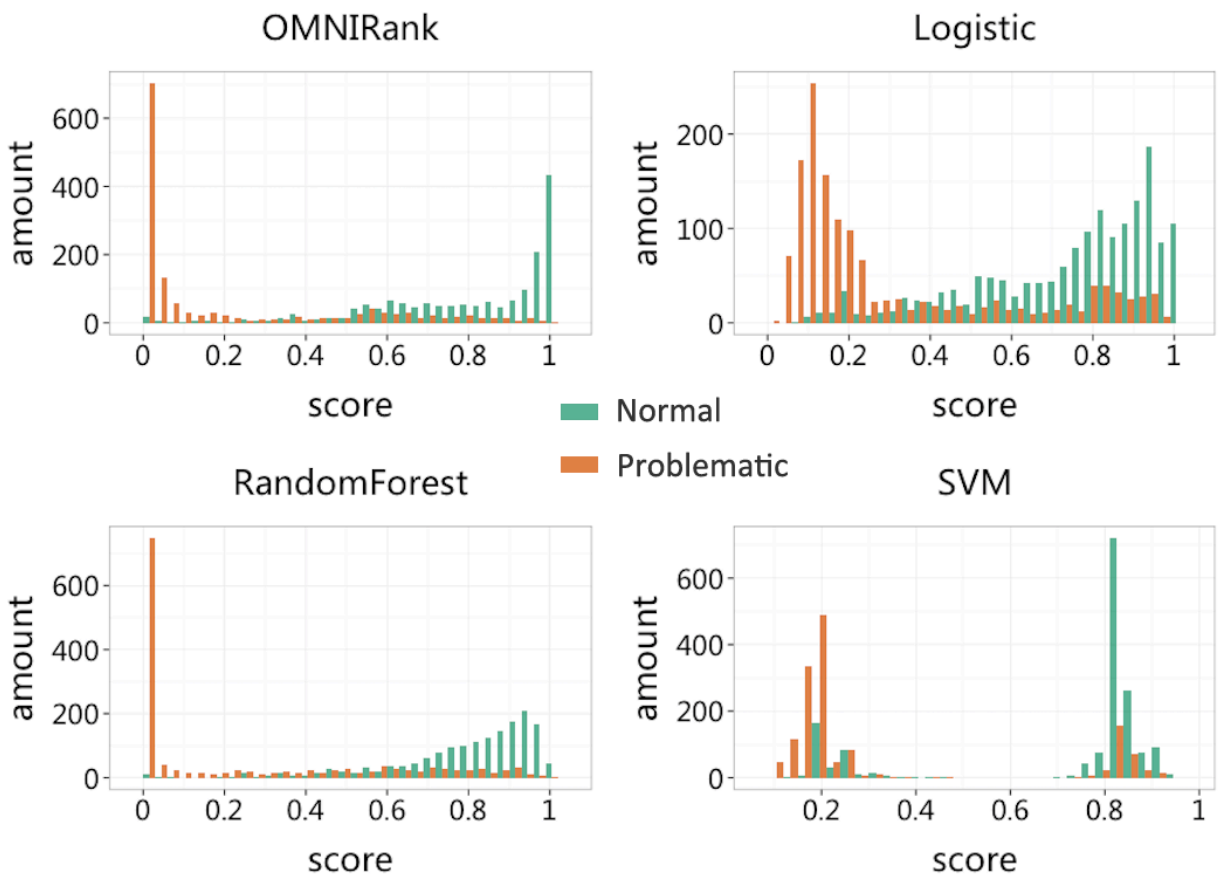

Fig.5 Score distributions of normal and problematic platforms

图 5 正常平台和问题平台的评分分布

在评估 OMNIRank 的分类性能时，是通过比较标记结果和当下时间的平台标签来计算模型的分类准确率。根据性质 3，为了评估 OMNIRank 的预测性能，本文以下一时间的平台标签为基准，统计当下时间评分排名不同区间中将要出现问题的平台比例，例如，对于 2016 年 3 月的评分排名，根据 2016 年 4 月的平台标签进行统计，以评估 OMNIRank 的预测性能。统计结果如表 4 所示，可以看出六个月以来，评分排名前 100 中没有出现任何问题平台，前 200 名中也只有在 2015 年 12 月出现了 1 家问题平台，说明对投资者而言，OMNIRank 风险量化排名前 200 是一个相对安全可靠的投资区间。除此之外还可以发现，随着排名区间范围的扩大，区间所含平台的平均利率也逐渐提高，说明高回报确实伴随着高风险，这为资深投资者

Table 4 Evaluations of prediction performance of OMNIRank

表 4 OMNIRank 预测性能评估

| 排名区间 | 2015.11 | 2015.12 | 2016.01 | 2016.02 | 2016.03 | 2016.04 | 平均利率（%） |
|---|---|---|---|---|---|---|---|
| 前 20 | 0 | 0 | 0 | 0 | 0 | 0 | 10.71 |
| 前 50 | 0 | 0 | 0 | 0 | 0 | 0 | 10.99 |
| 前 100 | 0 | 0 | 0 | 0 | 0 | 0 | 11.42 |
| 前 200 | 0 | 0.5 | 0 | 0 | 0 | 0 | 12.64 |
| 前 500 | 0.4 | 1.0 | 1.0 | 0 | 0.6 | 0.2 | 13.58 |
| 前 1000 | 0.5 | 2.1 | 2.1 | 1.2 | 2.2 | 1.0 | 14.14 |
| 全部 | 2.71 | 4.85 | 4.39 | 4.08 | 4.31 | 1.54 | 14.51 |

如何综合风险和利率以获得最大期望回报提供了有力的数据支持。

## 6 数据可视化

本文根据 OMNIRank 的风险评分排名，筛选出前 100 名平台，并基于其对应数据进行了丰富的数据可视化和交互探索[4]，以便投资者结合自身实际投资偏好选择适合自己的 P2P 平台，所使用的数据可视化方法和形式包括折线图、饼图、柱状图、散点图、雷达图、矩形树图、字符云、平行坐标轴、平行时间流、新闻时间流、知识图谱、演化时间线等[10]。

数据可视化部分设计了四个页面，行业概览、平台详情、平台对比、懒人选投，如图 6～7 所示。通过行业概览可以了解 P2P 行业整体的发展趋势和统计数据，如平台知识图谱、地理分布、地域统计、指数对比等；通过平台详情可以查看 P2P 平台的详细信息，如注册信息、人员信息、核心指标、舆情统计、历史数据、大事时间线等；在平台对比中可以选择任意两家 P2P 平台进行全面的对比，懒人选投则设计了基于选择的平台推荐、基于标签的平台匹配、基于图谱的平台关联三项功能。通过以上功能，投资者可以首先选择是否要进入 P2P 网贷这一行业，然后选择一些感兴趣的平台了解详情，通过深入对比进一步决策取舍，并借鉴智能推荐作出最后的投资选择。

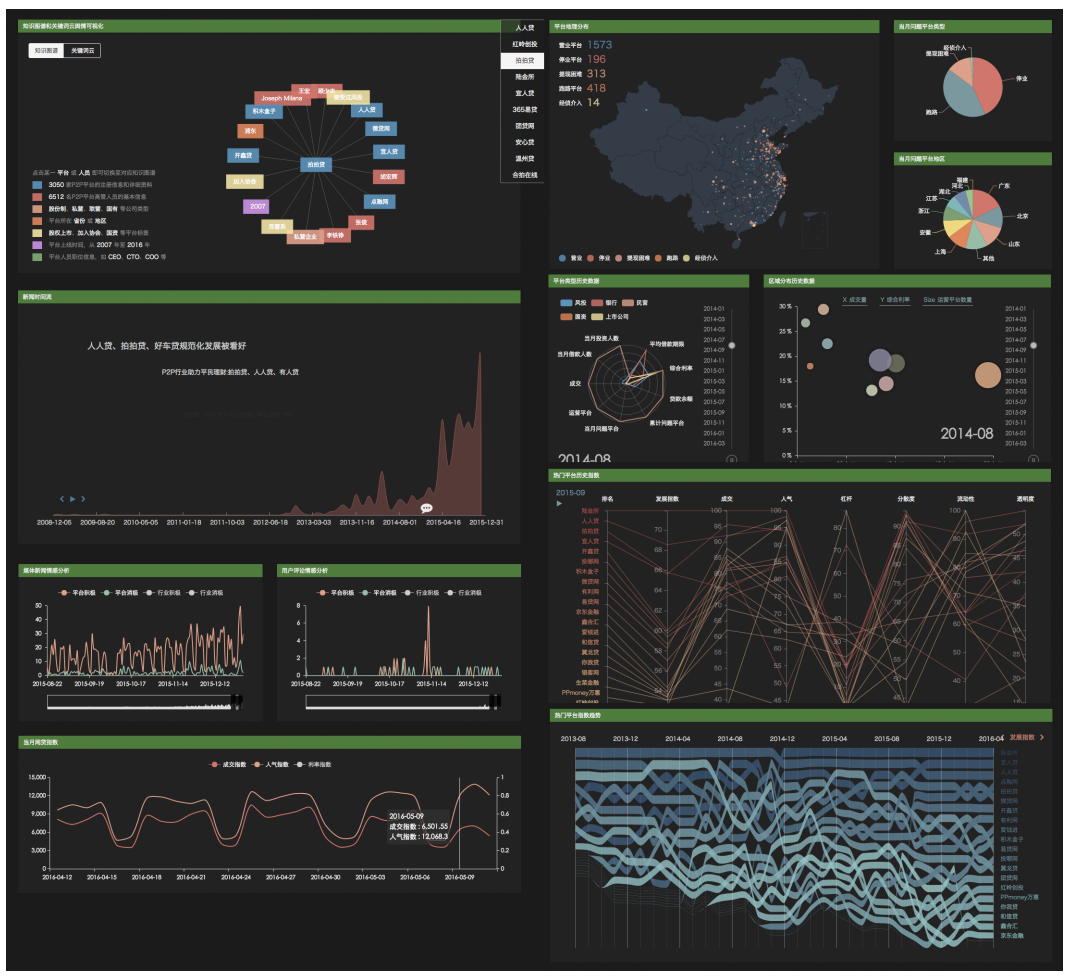

Fig.6 Business overview

图 6 行业概览

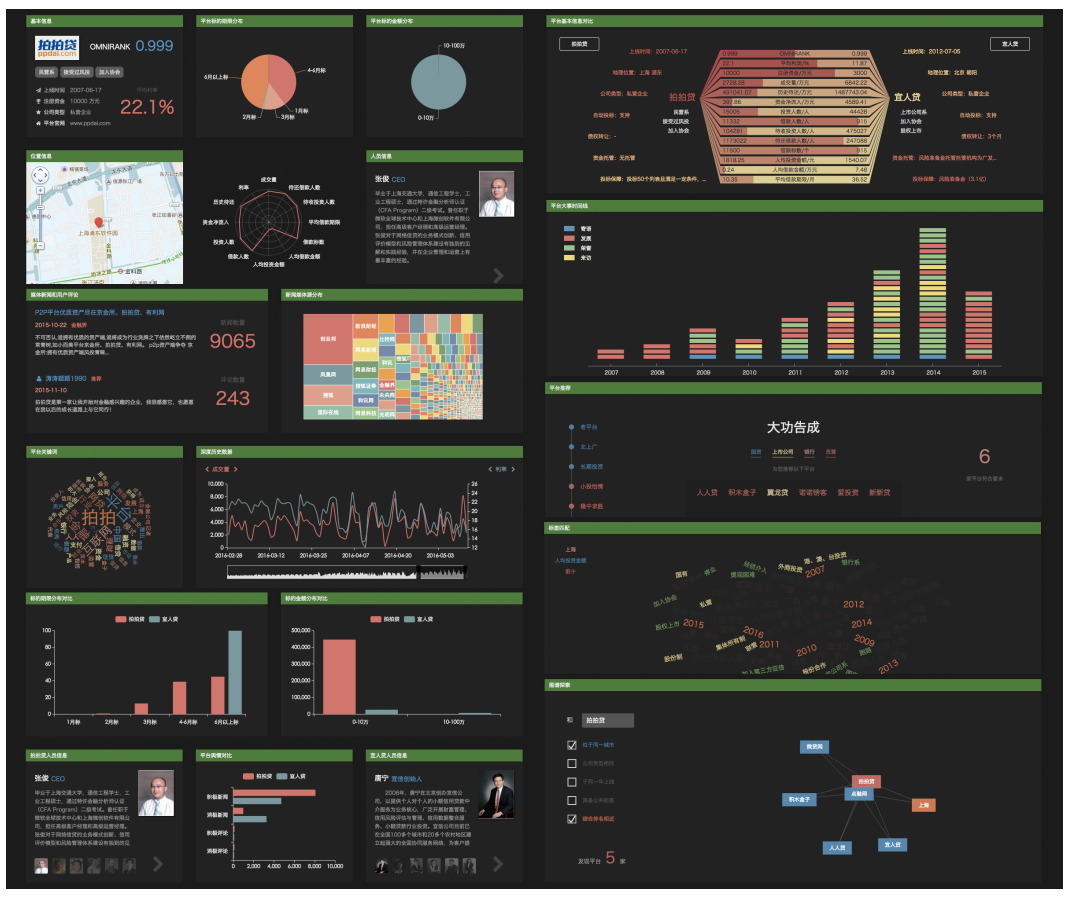

Fig.7 Platform details, Platform comparison, Recommendations

图 7 平台详情、平台对比、懒人选投

## 7 结束语

本文针对国内愈发严重的 P2P 平台倒闭问题，提出了基于深度学习的平台风险量化模型 OMNIRank，在获取和清洗了相关的海量多源异构数据之后，采用数据分析提取了平台风险的多维特征，通过多种深度学习模型组成的复杂网络进行风险量化。OMNIRank 实现了优越的分类性能和预测性能，结合丰富的数据可视化和交互探索，能够为投资者辨别问题平台提供有力的理论指导和实际价值。

4 在线可视化链接：http://zhanghonglun.cn/ppd/

# 参考文献

## 附录

1 全部数据集：http://data.sjtu.edu.cn/dataset/ppd-stay-foolish

2 OMNIRank：https://github.com/wang-haiyang/ppd_model

3 可视化代码：https://github.com/Honlan/ppd-magic-mirror

4 可视化链接：http://zhanghonglun.cn/ppd